\documentclass[twocolumn,english,superscriptaddress,citeautoscript,showpacs,preprintnumbers,amsmath,amssymb,prb,floatfix,footinbib]{revtex4}
\usepackage[utf8x]{inputenc}
\usepackage[scaled=2]{helvet}
\usepackage[T1]{fontenc}
\usepackage[utf8]{luainputenc}
\usepackage{multirow}
\usepackage{textcomp}
\usepackage{amsmath}
\usepackage{amssymb}
\usepackage{graphicx}
\usepackage{subscript}
\usepackage{epstopdf}

\makeatletter

\@ifundefined{textcolor}{}
{%
 \definecolor{BLACK}{gray}{0}
 \definecolor{WHITE}{gray}{1}
 \definecolor{RED}{rgb}{1,0,0}
 \definecolor{GREEN}{rgb}{0,1,0}
 \definecolor{BLUE}{rgb}{0,0,1}
 \definecolor{CYAN}{cmyk}{1,0,0,0}
 \definecolor{MAGENTA}{cmyk}{0,1,0,0}
 \definecolor{YELLOW}{cmyk}{0,0,1,0}
}

\usepackage[scaled=2]{helvet}\usepackage{amstext}

\makeatletter
\@ifundefined{textcolor}{}{\definecolor{BLACK}{gray}{0}\definecolor{WHITE}{gray}{1}\definecolor{RED}{rgb}{1,0,0}\definecolor{GREEN}{rgb}{0,1,0}\definecolor{BLUE}{rgb}{0,0,1}\definecolor{CYAN}{cmyk}{1,0,0,0}\definecolor{MAGENTA}{cmyk}{0,1,0,0}\definecolor{YELLOW}{cmyk}{0,0,1,0}}

\usepackage{dcolumn}
\usepackage{bm}
\usepackage{times}
\usepackage{hyperref}
\hypersetup{colorlinks=true,linkcolor=red,citecolor=blue}
\makeatother

\usepackage{babel}

\makeatother

\usepackage{babel}
\begin{document}

\title{Hydrostatic pressure effects on the static magnetism in Eu(Fe$_{0.925}$Co$_{0.075}$)$_{2}$As$_{2}$}

\author{W. T. Jin}

\email{w.jin@fz-juelich.de}

\affiliation{J\"{u}lich Centre for Neutron Science JCNS at Heinz Maier-Leibnitz Zentrum (MLZ), Forschungszentrum J\"{u}lich GmbH, Lichtenbergstraße 1, D-85747 Garching, Germany}

\author{J. -P. Sun}

\affiliation{Beijing National Laboratory for Condensed Matter Physics and Institute of Physics, Chinese Academy of Sciences, Beijing 100190, China}

\author{G. Z. Ye}

\affiliation{Beijing National Laboratory for Condensed Matter Physics and Institute of Physics, Chinese Academy of Sciences, Beijing 100190, China}

\author{Y. Xiao}

\affiliation{J\"{u}lich Centre for Neutron Science JCNS and Peter Gr\"{u}nberg Institut PGI, JARA-FIT, Forschungszentrum J\"{u}lich GmbH, D-52425 J\"{u}lich, Germany}

\author{Y. Su}

\affiliation{J\"{u}lich Centre for Neutron Science JCNS at Heinz Maier-Leibnitz Zentrum (MLZ), Forschungszentrum J\"{u}lich GmbH, Lichtenbergstraße 1, D-85747 Garching, Germany}

\author{K. Schmalzl}

\affiliation{J\"{u}lich Centre for Neutron Science JCNS at Institut Laue-Langevin (ILL), Forschungszentrum J\"{u}lich GmbH, Boite Postale 156, 38042 Grenoble Cedex 9, France}

\author{S. Nandi}

\affiliation{Department of Physics, Indian Institute of Technology, Kanpur 208016, India}

\author{Z. Bukowski }

\affiliation{Institute of Low Temperature and Structure Research, Polish Academy of Sciences, 50-422 Wroclaw, Poland}

\author{Z. Guguchia}

\affiliation{Department of Physics, Columbia University, New York, NY 10027, USA}

\author{E. Feng}

\affiliation{J\"{u}lich Centre for Neutron Science JCNS at Heinz Maier-Leibnitz Zentrum (MLZ), Forschungszentrum J\"{u}lich GmbH, Lichtenbergstraße 1, D-85747 Garching, Germany}

\author{Z. Fu}

\affiliation{J\"{u}lich Centre for Neutron Science JCNS at Heinz Maier-Leibnitz Zentrum (MLZ), Forschungszentrum J\"{u}lich GmbH, Lichtenbergstraße 1, D-85747 Garching, Germany}

\author{J. -G. Cheng}

\email{jgcheng@iphy.ac.cn}

\affiliation{Beijing National Laboratory for Condensed Matter Physics and Institute of Physics, Chinese Academy of Sciences, Beijing 100190, China}

\begin{abstract}
The effects of hydrostatic pressure on the static magnetism in Eu(Fe$_{0.925}$Co$_{0.075}$)$_{2}$As$_{2}$ are investigated by complementary electrical resistivity, ac magnetic susceptibility and single-crystal neutron diffraction measurements. A specific pressure-temperature phase diagram of Eu(Fe$_{0.925}$Co$_{0.075}$)$_{2}$As$_{2}$ is established. The structural phase transition, as well as the spin-density-wave order of Fe sublattice, is suppressed gradually with increasing pressure and disappears completely above 2.0 GPa. In contrast, the magnetic order of Eu sublattice persists over the whole investigated pressure range up to 14 GPa, yet displaying a non-monotonic variation with pressure. With the increase of the hydrostatic pressure, the magnetic state of Eu evolves from the canted antiferromagnetic structure in the ground state, via a pure ferromagnetic structure under the intermediate pressure, finally to a possible "novel'' antiferromagnetic structure under the high pressure. The strong ferromagnetism of Eu coexists with the pressure-induced superconductivity around 2 GPa. The change of the magnetic state of Eu in Eu(Fe$_{0.925}$Co$_{0.075}$)$_{2}$As$_{2}$ upon the application of hydrostatic pressure probably arises from the modification of the indirect Ruderman-Kittel-Kasuya-Yosida (RKKY) interaction between the Eu$^{2+}$ moments tuned by external pressure.
\end{abstract}
\maketitle

\section{Introduction}

The discovery of Fe-based superconductors \cite{Kamihara_08} has provided new platforms to study the intriguing interplay between superconductivity (SC) and magnetism. SC in these novel materials was found to be in close proximity to the magnetism, as it emerges when the long-range antiferromagnetic (AFM) order in the parent compounds gets well suppressed by means of chemical doping or the application of external pressure,\cite{Johnston_10} and the spin fluctuations are believed to be responsible for the unconventional SC in them.\cite{Paglione_10,Dai_15} 

Among various classes of Fe-based superconductors, the EuFe$_{2}$As$_{2}$-based compounds (Eu-122) have drawn tremendous attention, as they contain two magnetic sublattices and show strong spin-charge-lattice coupling. \cite{Xiao_10,Xiao_12} In a purely ionic picture, the $\mathit{S}$-state (orbital moment $\mathit{L}$ = 0) Eu$^{2+}$ rare-earth ion has a 4$\mathit{f}$$^{7}$ electronic configuration and a total electron spin $\mathit{S}$ = 7/2, corresponding to a theoretical effective magnetic moment of 7.94 $\mathit{\mu_{B}}$.\cite{Marchand_78} EuFe$_{2}$As$_{2}$ undergoes a structural phase transition from a tetragonal to an orthorhombic phase at 190 K, accompanied by a spin-density-wave (SDW) order of the itinerant Fe moments. In addition, the localized Eu$^{2+}$ spins order below 19 K in the A-type AFM structure (ferromagnetic layers stacked antiferromagnetically along the $\mathit{c}$ axis).\cite{Jiang_09_NJP,Herrero-Martin_09,Xiao_09} The undoped parent compound EuFe$_{2}$As$_{2}$ can be tuned into a superconductor by chemical substitutions into the Eu-,\cite{Jeevan_08} Fe-,\cite{Jiang_09,Jiao_11,Jiao_13} or As-site,\cite{Ren_09} respectively. The SC can also be realized by the application of external physical pressure in undoped EuFe$_{2}$As$_{2}$ with the superconducting transition temperature $\mathit{T_{sc}}$ $\sim$ 30 K in a narrow range of 2.5-3.0 GPa.\cite{Miclea_09,Terashima_09,Matsubayshi_11} 

Recently, considerable experimental efforts have been devoted to understand how the magnetism in both sublattices develops with different chemical doping.\cite{Anupam_11,Zhang_12,Jiang_09,Blachowski_11,Guguchia_11_NMR,Guguchia_11,Jin_EuCoPhaseDiagram,Cao_11,Jeevan_11,Zapf_11,Guguchia_13,Zapf_13,Jiao_12,Jin_16}
It is well established that the SDW transition of the Fe sublattice gets gradually suppressed with increasing doping level, while the magnetic order of the Eu sublattice persists over the whole doping region. The magnetic ground state of the Eu$^{2+}$ moments displays
a systematic change with increasing doping concentration, from the A-type AFM structure (with the spins lying in the $\mathit{ab}$ planes) at low doping levels to the pure ferromagnetic structure (with the spins aligning along the $\mathit{c}$ axis) at high doping levels.\cite{Jin_EuCoPhaseDiagram} Interestingly, the strong ferromagnetic (FM) order of the localized Eu$^{2+}$ spins, with a huge moment close to 7 $\mu_{B}$ per Eu, was confirmed to be compatible with the SC.\cite{Ren_09,Jiao_11,Jin_13,Jiao_13,Nandi_14,Nandi_14_neutron,Jin_15,Anand_15} 

Nevertheless, to the best of our knowledge, only a limited number of studies about the pressure effects on the Eu magnetism in the Eu-122 compounds exist. For the undoped parent compound EuFe$_{2}$As$_{2}$, the high-pressure ac magnetic susceptibility measurement using the piston-cylinder cell suggests that the magnetic ground state of the Eu$^{2+}$ moments is still an AFM order in the pressure-induced superconducting phase (with the maximum applied pressure $\mathit{P}$ \textasciitilde{} 2.8 GPa),\cite{Terashima_09} similar to that under the ambient pressure. Further measurements using a cubic anvil cell indicate that the AFM order of the Eu$^{2+}$ moments persists up to an applied pressure $\mathit{P}$ \textasciitilde{} 6 GPa, above which it changes to the FM order, as confirmed by high-pressure x-ray magnetic circular dichroism (XMCD) experiments.\cite{Matsubayshi_11} The SDW transition of Fe gets completely suppressed at the critical pressure $\mathit{P_{C}}$ where the SC emerges. In addition, complementary high-pressure muon-spin rotation ($\mu$SR) and magnetization measurements were performed on non-superconducting isovalent-substituted EuFe$_{2}$(As$_{0.88}$P$_{0.12}$)$_{2}$, in which the Eu$^{2+}$ spins were found to order in the canted AFM (C-AFM) structure in the ground state.\cite{Guguchia_13} Possible superconducting phase (``X'' phase as referred in Ref.\onlinecite{Guguchia_13} ) was realized in EuFe$_{2}$(As$_{0.88}$P$_{0.12}$)$_{2}$ within a very narrow pressure range of 0.36-0.5 GPa, coexisting with the magnetic order of both the Eu and Fe moments. However, the magnetic structure of Eu in the so-called ``X'' phase can not be unambiguously determined there. 

In order to conclude how the magnetism in both sublattices develop with the external pressure and to clarify the nature of the magnetic state in possible pressure-induced superconducting phase, we have carried out complementary experiments including the electrical resistivity,
ac magnetic susceptibility and neutron diffraction measurements on the Eu(Fe$_{1-x}$Co$_{x}$)$_{2}$As$_{2}$ ($\mathit{x}$ = 0.075) single crystal under hydrostatic pressure. There are two reasons of choosing Eu(Fe$_{0.925}$Co$_{0.075}$)$_{2}$As$_{2}$ for the high-pressure studies. Firstly, according to the established $\mathit{T-x}$ phase diagram of electron-doped Eu(Fe$_{1-x}$Co$_{x}$)$_{2}$As$_{2}$,\cite{Jin_EuCoPhaseDiagram} the sample with $\mathit{x}$ = 0.075 is close to the superconducting dome. The superconducting phase might be reachable by applying moderate hydrostatic pressure. Secondly, the magnetic ground-state of the Eu sublattice in Eu(Fe$_{0.925}$Co$_{0.075}$)$_{2}$As$_{2}$ has been determined to be a canted-AFM structure.\cite{Jin_EuCoPhaseDiagram} The Eu$^{2+}$ spins are canted out of the $\mathit{ab}$ planes with an angle of 23.8(6)$^{\circ}$, giving rise to a net ferromagnetic moment component along the $\mathit{c}$ axis. It is thus very interesting to investigate how this intermediate magnetic structure in the $\mathit{T-x}$ phase diagram evolves with hydrostatic pressure, and to conclude what type of magnetic order of Eu can coexist with the pressure-induced superconducting phase.

\section{Experimental Details}

The single crystal of Eu(Fe$_{0.925}$Co$_{0.075}$)$_{2}$As$_{2}$
was grown out of the Sn flux.\cite{Jin_EuCoPhaseDiagram} The concentration of Co was determined by wavelength dispersive spectroscopy (WDS). Both the ambient-pressure and high-pressure neutron diffraction experiments were performed on the thermal-neutron two-axis diffractometer D23 at the Institut Laue Langevin (Grenoble, France). A Cu (2 0 0) monochromator was chosen to produce a monochromatic neutron beam with the wavelength of 1.279 Å. For the ambient-pressure measurement, a 76 mg platelike single crystal with dimensions $\sim$ 6 \texttimes{} 5 \texttimes{} 1 mm$^{3}$ was mounted on a thin Al plate with a small amount of GE varnish, and put inside a standard orange cryostat. For the high-pressure measurement, a 10 mg rectangular strip with dimensions $\sim$ 4 \texttimes{} 1 \texttimes{} 1 mm$^{3}$ was cut from the same piece of crystal, and put inside a TiZr gasket together with some lead powders as the pressure medium. The gasket was then mounted into the VX-5 type Paris-Edinburgh pressure cell \cite{Klotz_96} loaded with He gas for low-temperature
measurements in a 4 K dedicated cryostat. The pressure values were determined from the equation of state of lead,\cite{Strassle_14} based on the lattice parameters of lead measured by neutron diffraction at a certain temperature. For both experimental conditions, the crystals
were oriented with the orthorhombic $\mathit{b}$ axis (or $\mathit{a}$ axis due to twinning) lying vertical, so that the ($\mathit{H}$ 0 $\mathit{L}$) scattering plane can be accessible horizontally. (The orthorhombic notation will be used throughout this paper for convenience.)

High-pressure resistivity and ac magnetic susceptibility were measured in the Institute of Physics, Chinese Academy of Sciences, by using a self-clamped piston-cylinder cell (PCC) up to 2.2 GPa and a \textquotedblleft{}Palm\textquotedblright{} cubic anvil cell (CAC) up to 14.2 GPa. The standard four-probe method was used for the resistivity measurements and the mutual induction method for the ac magnetic susceptibility measurements. The pressure inside the PCC was determined by monitoring the superconducting transition temperature of tin (Sn), which was placed together with the sample in the Teflon cell filled with the Daphne 7373 as the pressure transmitting medium (PTM). The pressure inside the CAC was calibrated at room temperature by observing the characteristic phase transitions of Bismuth (Bi). In this case, glycerol was used as the PTM.

\section{Experimental Results }

Figure 1 shows the temperature dependencies of the electrical resistivity,
$\rho(T)$, of the Eu(Fe$_{0.925}$Co$_{0.075}$)$_{2}$As$_{2}$
single crystal measured with the PCC (a) and the CAC (b), respectively.
At ambient pressure, an upturn in $\rho(T)$ appears at $\mathit{T_{S}}$
$\sim$ 152 K, corresponding to the structural phase transition, as
confirmed by neutron diffraction presented below. Here $\mathit{T_{S}}$
is defined as the minimum in the first derivative of $\rho(T)$, $d\rho(T)/dT$.
In addition, $\rho(T)$ shows another kink at $\mathit{T_{Eu}}$=
17 K, due to the magnetic order of the Eu$^{2+}$ moments.\cite{Footnote}
With increasing pressure,$\mathit{T_{S}}$ shifts gradually to the
lower temperature, as shown in the inset of Fig. 1(a). Above 2.3 GPa,
no upturn in $\rho(T)$ can be observed anymore (Fig. 1(b)), indicating
that the structural and SDW transitions get completely suppressed
at this pressure. However, $\mathit{T_{Eu}}$, the magnetic transition
temperature of Eu, seems quite insensitive to the applied pressure
and stays lower than 20 K for $\mathit{P}$ $\leq$2 GPa. As shown
in the inset of Fig. 1(b), with further increasing pressure, $\mathit{T_{Eu}}$
starts to change significantly, as revealed by the non-monotonous
change of the minimum in the second derivative of $\rho(T)$, $d^{2}\rho(T)/dT^{2}$.
After reaching a maximum value of $\sim$ 52 K at 6.8 GPa, $\mathit{T_{Eu}}$
decreases slightly with increasing pressure, to $\sim$ 41 K at 11.2
GPa. Interestingly, $\mathit{T_{Eu}}$ reverses to increase again
when more pressure is applied. At the maximum applied pressure of
14.2 GPa, $\mathit{T_{Eu}}$ reaches another maximum around 49 K.
It is worth noting that at 2.17 GPa, $\rho(T)$ shows a sharp drop
at $\mathit{T_{SC}}$ $\sim$ 25 K, suggesting the appearance of the
pressure-induced SC as reported previously in the parent compound.\cite{Miclea_09}
The superconducting nature at 2.17 GPa is also reflected in the ac
magnetic susceptibility data presented below. However, the transition
to a zero-resistivity state is hindered by the magnetic order of Eu,
as shown by another anomaly in $\rho(T)$ around 21 K. The pressure-induced
reentrant resistivity below the superconducting transition here resembles
that observed at ambient pressure in the Eu(Fe$_{0.89}$Co$_{0.11}$)$_{2}$As$_{2}$
single crystals grown from self-flux method,\cite{Jiang_09} ascribing
to the competition between the SC and the Eu magnetic order. 

\begin{figure}
\centering{}\includegraphics{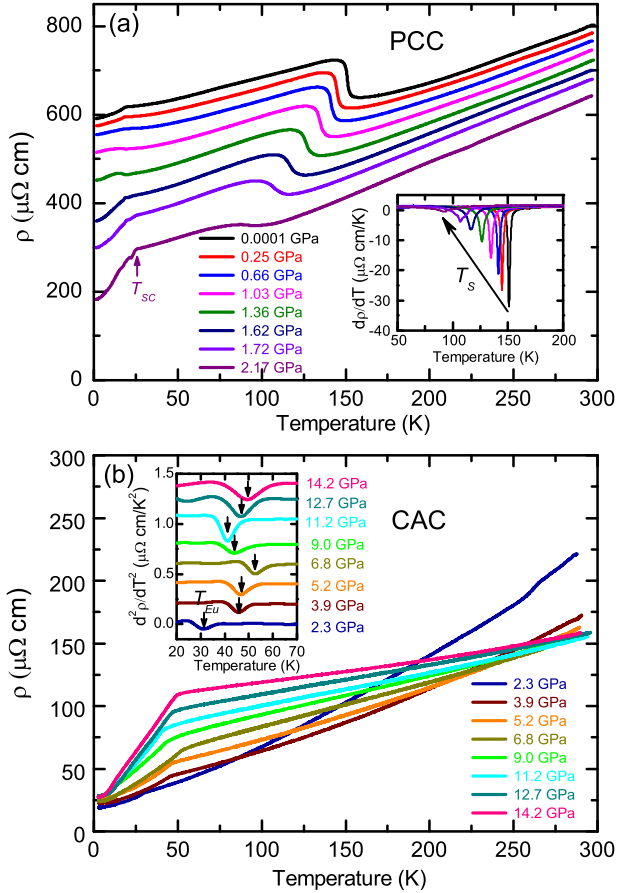}

\caption{Temperature dependencies of the electrical resistivity ($\rho)$ of
the Eu(Fe$_{0.925}$Co$_{0.075}$)$_{2}$As$_{2}$ single crystals
measured under high pressures with the piston-cylinder cell (a) and
the cubic anvil cell (b), respectively. }
\end{figure}

The ac magnetic susceptibility of the Eu(Fe$_{0.925}$Co$_{0.075}$)$_{2}$As$_{2}$
single crystal measured under high pressure with the PCC and the CAC
is shown in Fig. 2(a) and 2(b), respectively. At ambient pressure,
a peak at $\mathit{T_{Eu}}$ = 17 K in $\chi^{'}$, the real part
of the ac magnetic susceptibility, indicates the antiferromagnetic
order of the Eu$^{2+}$ spins. As determined from ambient-pressure
neutron diffraction, the magnetic ground state of the Eu$^{2+}$ moments
is a canted AFM structure with a net FM moment component along the
$\mathit{c}$ axis.\cite{Jin_EuCoPhaseDiagram} With increasing pressure,
$\mathit{T_{Eu}}$ shifted slightly to a lower temperature for $\mathit{P}$
$\leq$ 1.03 GPa, as shown in Fig. 2(a), suggesting that the interlayer
AFM coupling weakens under pressure. At 1.37 GPa, the antiferromagnetic
peak almost gets smeared out and instead a pronounced kink is observed
at $\mathit{T_{Eu}}$ = 18 K. The kink temperature shifts continuously
to a higher temperature with further increasing pressure, reaching
a maximum value of $\sim$ 52 K at 6.8 GPa (Fig. 2(c)). The distinct
tendencies of the evolution of $\mathit{T_{Eu}}$ with increasing
pressure for $\mathit{P}$ $\leq$ 1.03 GPa and 1.37 GPa $\leq$$\mathit{P}$
$\leq$ 6.8 GPa suggests that the magnetic structures of the Eu$^{2+}$
spins might be different in the two pressure regions. With further
increasing pressure ($\mathit{P}$ $\geq$ 6.8 GPa), $\mathit{T_{Eu}}$
decreases again, as shown in Fig. 2(c), to $\sim$ 43 K at 11.2 GPa.
The pressure dependence of $\mathit{T_{Eu}}$ for Eu(Fe$_{0.925}$Co$_{0.075}$)$_{2}$As$_{2}$
obtained from the ac magnetic susceptibility measurement is well consistent
with that extracted from the resistivity measurement. The anomaly
at $\mathit{T_{Eu}}$ can be hardly resolved in $\chi^{'}$ for $\mathit{P}$
$\geq$ 13.2 GPa, implying that the Eu$^{2+}$ moments probably order
antiferromagnetically again. Furthermore, $\chi^{'}$ shows additional
features for the pressures around 2.17 GPa. Compared with other pressures,
the $\chi^{'}(T)$ curves at 2.17 GPa exhibits some diamagnetic response
associated with superconductivity. As shown in Fig. 2(a), at $\mathit{P}$
= 2.17 GPa, the slope of the $\chi^{'}(T)$ curve shows a pronounced
downward bending around 12 K, which is most likely the result of the
competition between the magnetism of Eu and the pressure-induced superconductivity,
as hinted by the resistivity data at this pressure value. 

\begin{figure}
\centering{}\includegraphics{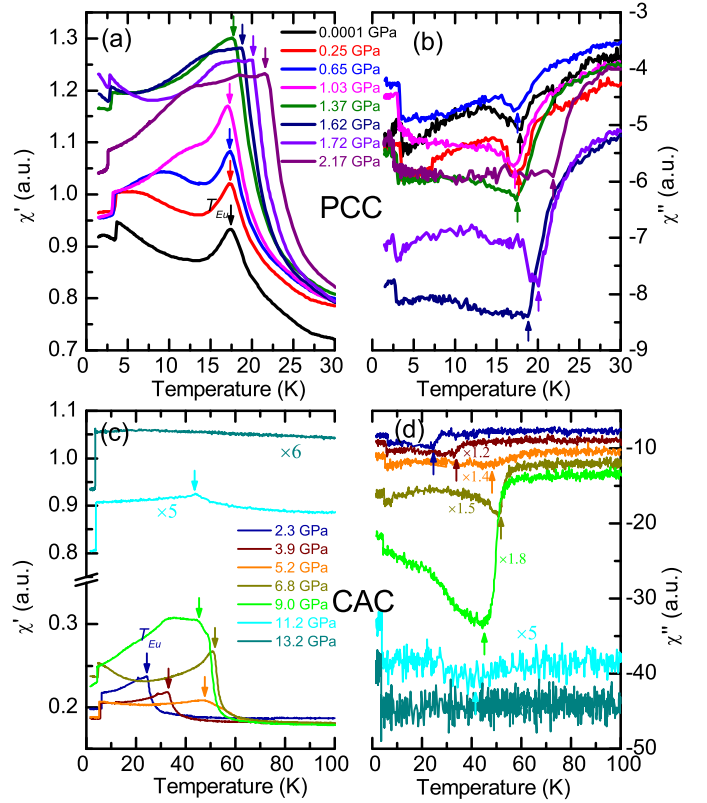}

\caption{Temperature dependencies of the real and imaginary part of the ac
magnetic susceptibility,$\chi'$ and $\chi''$ , of the Eu(Fe$_{0.925}$Co$_{0.075}$)$_{2}$As$_{2}$
single crystals measured under high pressures with the piston-cylinder
cell (a, b) and the cubic anvil cell (c, d), respectively. }
\end{figure}

In order to confirm the nature of the anomalies revealed by the macroscopic
measurements, the neutron diffraction measurements were performed
on the Eu(Fe$_{0.925}$Co$_{0.075}$)$_{2}$As$_{2}$ single crystal
at ambient pressure and under high pressure, respectively. Figure
3(a) shows the ambient-pressure temperature dependencies of the integrated
intensity of the (4 0 0) and (1 0 3) peaks, one strong nuclear reflection
of the orthorhombic phase and one magnetic reflection due to the SDW
order of the itinerant Fe moments, respectively. The rapid increase
of the intensity of the nulear (4 0 0) peak below $\mathit{T_{S}}$
= 151(1) K indicates the structural phase transition from a tetragonal
to an orthorhombic phase, as the emergence of the orthorhombic domains
has a strong impact on the extinction conditions of the nuclear Bragg
reflections. The transition temperature determined here is well consistent
with that determined from the resistivity measurement. In addition,
fitting to the intensity of the (1 0 3) reflection for the temperature
close to the transition yields the onset temperature of the SDW order
of Fe, $\mathit{T_{SDW}}$ = 148(1) K. Compared with the parent compound
EuFe$_{2}$As$_{2}$, both transitions are significantly suppressed
by 7.5\% Co doping. The size of the Fe$^{2+}$ moment is estimated
to be 0.63(4) $\mu_{B}$. Furthermore, at ambient pressure, the magnetic
ground state of Eu was determined to be a canted AFM structure with
a net FM moment component along the $\mathit{c}$ axis, as reported
in Ref.\onlinecite{Jin_EuCoPhaseDiagram} and illustrated as the inset
in Fig. 3(b). The Eu$^{2+}$ spins were found to be canted with an
angle of 23.8(6)$^{\circ}$ out of the $\mathit{ab}$ plane with the
moment size of 6.22(3) $\mu_{B}$. The magnetic ordering temperature
of Eu was determined to be 17.0(2) K according to the temperature
dependencies of both the (2 0 0) and (0 0 3) magnetic peaks, as shown
in Fig. 3(b).

\begin{figure}
\centering{}\includegraphics{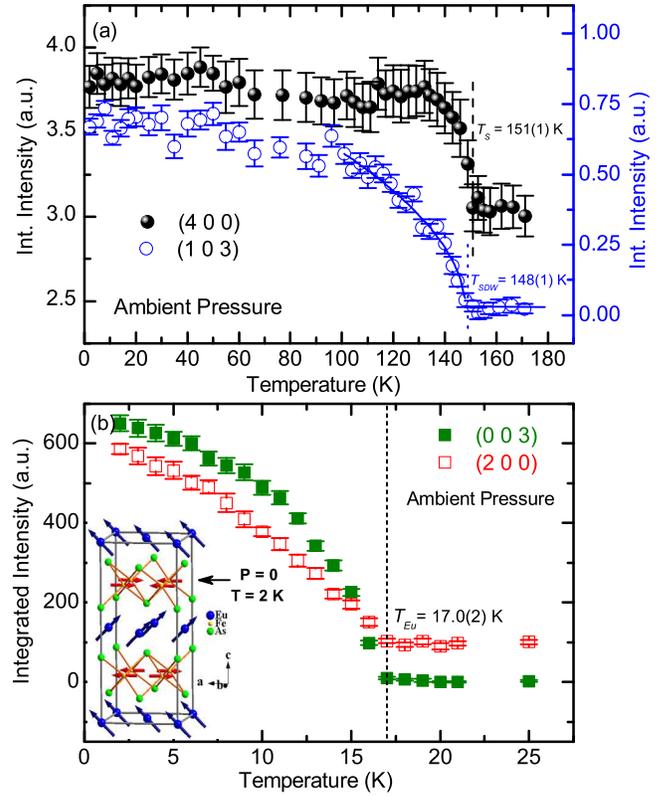}

\caption{Temperature dependencies of the integrated intensity of the (4 0 0)
and (1 0 3) reflections (a), and the (0 0 3) and (2 0 0) reflections
(b), respectively, measured by neutron diffraction on the Eu(Fe$_{0.925}$Co$_{0.075}$)$_{2}$As$_{2}$
single crystal at ambient pressure. The dash line and dot line in
(a) mark the structural phase transition and the SDW transition of
Fe, respectively. The short dash line in (b) marks the magnetic transition
associated with the canted AFM order of the Eu$^{2+}$ spins. The
ground-state magnetic structure of Eu(Fe$_{0.925}$Co$_{0.075}$)$_{2}$As$_{2}$
is illustrated as an inset in (b).\cite{Jin_EuCoPhaseDiagram} }
\end{figure}

Figure 4 shows the temperature dependencies of the integrated intensity
of the (4 0 0) nuclear peak measured under the high pressure at 2.0
GPa, 3.7 GPa and 6.6 GPa, respectively. Different from a rapid increase
below $\mathit{T_{S}}$ = 151(1) K at ambient pressure (see Fig. 3(a)),
the intensity of (4 0 0) remains almost constant for the temperature
range from 15 K to 160 K when a pressure of 2.0 GPa is applied, as
shown in the inset of Fig. 4, indicating the complete suppression
of the structural phase transition in Eu(Fe$_{0.925}$Co$_{0.075}$)$_{2}$As$_{2}$
under the pressure larger than 2.0 GPa ($\mathit{P}$ $\geq$ 2.0
GPa). This is slightly inconsistent with the result from the resistivity
measurements, since an upturn can still be resolved around 92 K in
$\rho(T)$ at 2.17 GPa. The discrepancy between two probes might be
due to the difference in the hydrostaticity of the pressure generated
by the piston-cylinder cell and the Paris-Edinburgh cell. As presented
above, the signature of superconductivity is exhibited in the macroscopic
measurements for the pressure close to 2.17 GPa. Therefore, the complete
suppression of the structural phase transition at 2.0 GPa is in line
with the expectation that the superconductivity emerges in close proximity
to the criticality where the structural distortion, as well as the
following or accompanying SDW order of Fe, disappears.

\begin{figure}
\centering{}\includegraphics{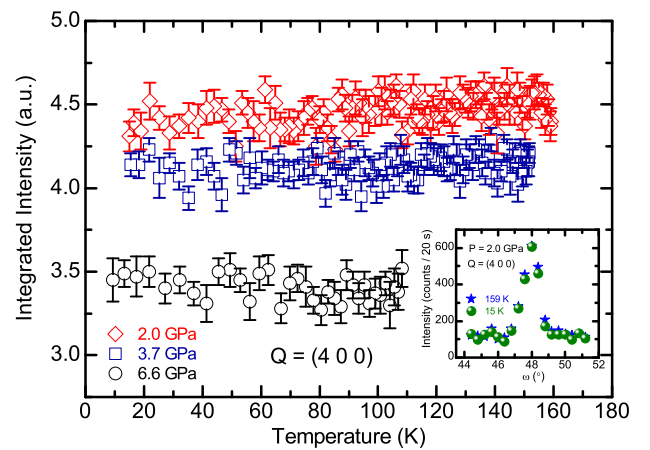}

\caption{Temperature dependencies of the integrated intensity of the (4 0 0)
nuclear reflection of the Eu(Fe$_{0.925}$Co$_{0.075}$)$_{2}$As$_{2}$
single crystal under the high pressure at 2.0 GPa, 3.7 GPa and 6.6
GPa, respectively. The rocking scans of the (4 0 0) peak at 159 K
and 15 K under the pressure at 2.0 GPa are shown in the inset.}
\end{figure}

\begin{figure*}
\centering{}\includegraphics[width=1\textwidth]{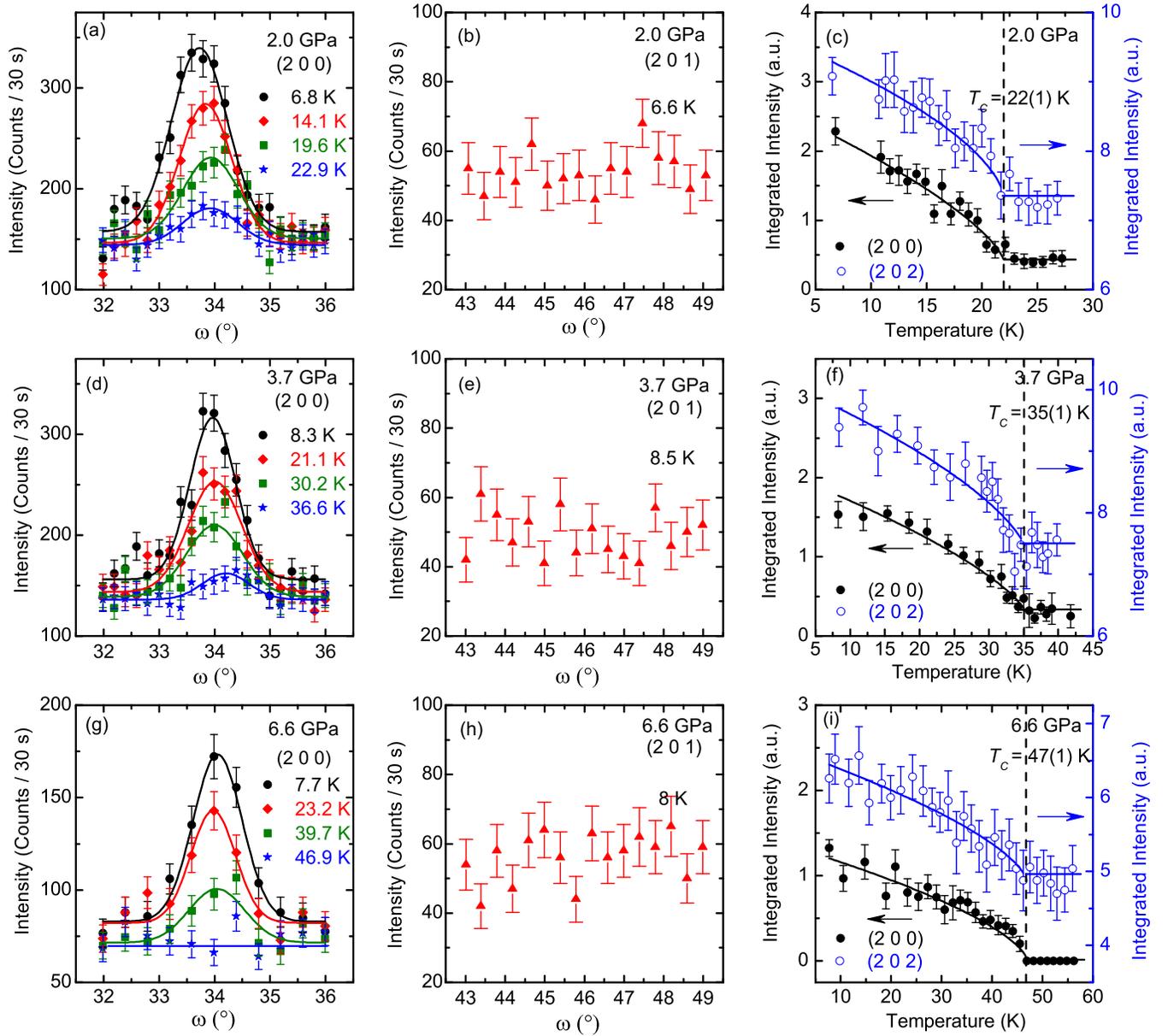}

\caption{The summary of the neutron diffraction data on the Eu(Fe$_{0.925}$Co$_{0.075}$)$_{2}$As$_{2}$
single crystal measured under the high pressure at 2.0 GPa (a-c),
3.7 GPa (d-f) and 6.6 GPa (g-i), respectively. The rocking scans of
the (2 0 0) and (2 0 1) reflections at different temperatures measured
at 2.0 GPa, 3.7 GPa, and 6.6 GPa are shown in (a, b), (d, e) and (g,
h), respectively. The solid curves represent the fits using the Gaussian
profiles. The integrated intensity of the (2 0 0) and (2 0 2) peaks
under different pressure values are plotted as functions of the temperature
in (c), (f), and (i), respectively. The vertical dashed lines mark
the ferromagnetic transition temperatures of Eu at different $\mathit{P}$
and the solid lines are guides to the eye.}
\end{figure*}
The effect of hydrostatic pressure on the magnetic reflections arising
from the magnetic order of Eu is summarized in Figure 5. As shown
in Figs. 5(a), 5(d) and 5(g), the intensity of the (2 0 0) peak always
weakens significantly with increasing temperature, indicating a strong
ferromagnetic contribution on this reflection, at 2.0, 3.7 and 6.6
GPa, respectively. On the other hand, the (2 0 1) reflection observed
at ambient pressure due to the antiferromagnetic interlayer coupling
of the Eu$^{2+}$ spins disappears upon the application of the pressure
larger than 2.0 GPa (Figs. 5(b), 5(e) and 5(h)), suggesting a pure
ferromagnetic order of the Eu$^{2+}$ moments along the $\mathit{c}$
axis at the applied pressure values. The temperature dependencies
of the integrated intensity of both the (2 0 0) and (2 0 2) reflections
as plotted in Figs. 5(c), 5(f) and 5(i) allow us to determine the
ferromagnetic transition temperature ($\mathit{T_{C}}$) as 22(1),
35(1) and 47(1) K, for 2.0, 3.7 and 6.6 GPa, respectively. The observation
here that $\mathit{T_{C}}$ shifts to a higher temperature with increasing
pressure is well consistent with the results from both the resistivity
and the ac magnetic susceptibility measurements within the same pressure
region. It is worth noting that the net ferromagnetic contribution
on the nuclear scattering part of the (2, 0, 2) reflection at the
lowest temperature is estimated to be around 24 \% for all the three
pressure values, further corroborating the ferromagnetic alignment
of the Eu$^{2+}$ moments completely along the $\mathit{c}$ axis
at 2.0, 3.7 and 6.6 GPa.\cite{Jin_13,Jin_EuCoPhaseDiagram}

\section{Discussion and Conclusion}

Combining the results from the resistivity, ac magnetic susceptibility
and neutron diffraction measurements, a phase diagram describing how
the static magnetism of Eu(Fe$_{0.925}$Co$_{0.075}$)$_{2}$As$_{2}$
develops upon the application of hydrostatic pressure is established.
As shown in Fig. 6, the structural phase transition (as well as the
SDW order of Fe) gets gradually suppressed with increasing pressure
and disappears at a critical pressure $\mathit{P_{c}}$ $\sim$ 2.0
GPa, which is lower than $\mathit{P_{c}}$ $\sim$ 2.5-2.7 GPa in
the parent compound EuFe$_{2}$As$_{2}$,\cite{Matsubayshi_11} ascribing
to additional electron-doping effect from Co.

\begin{figure}
\centering{}\includegraphics{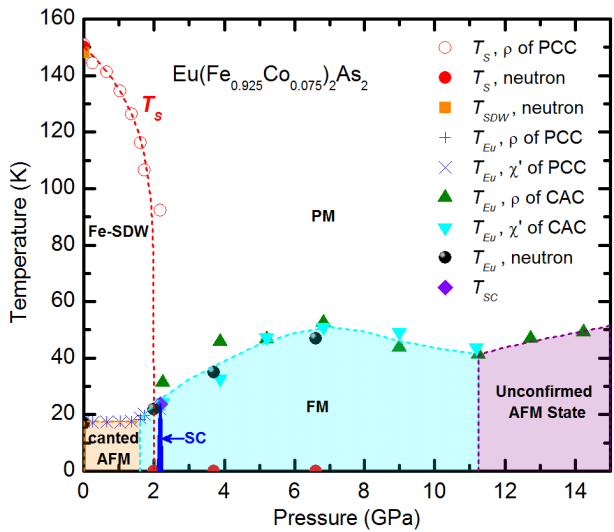}

\caption{Pressure-temperature phase diagram of Eu(Fe$_{0.925}$Co$_{0.075}$)$_{2}$As$_{2}$
determined from the resistivity, ac magnetic susceptibility, and neutron
diffraction measurements. }
\end{figure}

The magnetic order of Eu persists over the whole range of the applied
pressure up to 14 GPa, yet displaying a non-monotonic variation with
pressure. Below 1.5 GPa, $\mathit{T_{Eu}}$, the ordering temperature
of the Eu$^{2+}$ spins, stays almost constant, suggesting a canted
antiferromagnetic structure with a net ferromagnetic moment component
along the $\mathit{c}$ direction, as observed at the ambient pressure.
With further increasing pressure, $\mathit{T_{Eu}}$ starts to increase,
reaching a maximum value of $\sim$ 50 K at 7 GPa. The magnetic structure
of Eu in this pressure range is revealed by neutron diffraction to
be a pure ferromagnetic order along the $\mathit{c}$ axis. The role
of the hydrostatic pressure in driving the Eu$^{2+}$ moments to order
ferromagnetically in Eu(Fe$_{0.925}$Co$_{0.075}$)$_{2}$As$_{2}$
seems quite similar to the effect of introducing more electrons by
the means of further Co doping, as reflected by the ambient-pressure
phase diagram of Eu(Fe$_{1-x}$Co$_{x}$)$_{2}$As$_{2}$.\cite{Jin_EuCoPhaseDiagram}
Above 7 GPa, $\mathit{T_{Eu}}$ declines slightly with increasing
pressure to $\sim$ 40 K at 11.2 GPa, indicating the weakening of
the ferromagnetic coupling between the Eu$^{2+}$ moments in this
pressure range. The suppression of the FM state was also observed
in the parent compound EuFe$_{2}$As$_{2}$ for $\mathit{P}$ $\geq$
8 GPa, which was explained as the result of the valance change from
magnetic Eu$^{2+}$ to nonmagnetic Eu$^{3+}$ state as observed by
x-ray absorption spectroscopy (XAS) under high pressure.\cite{Matsubayshi_11}
The pressure dependence of $\mathit{T_{Eu}}$ in Eu(Fe$_{0.925}$Co$_{0.075}$)$_{2}$As$_{2}$
for $\mathit{P}$ $\leq$ 11 GPa is quite similar to that in the parent
compound EuFe$_{2}$As$_{2}$,\cite{Matsubayshi_11} showing an almost
constant value in the AFM or canted AFM region and a dome-like variation
in the FM region.

Interestingly, as revealed by the resistivity measurements,$\mathit{T_{Eu}}(P)$
reverses to increase again for Eu(Fe$_{0.925}$Co$_{0.075}$)$_{2}$As$_{2}$
when more pressure is applied, which was not observed for the parent
compound. As revealed by high-pressure XAS, the average valence state
of Eu in both EuFe$_{2}$As$_{2}$ and EuCo$_{2}$As$_{2}$ increases
with the applied pressure due to a part transition from Eu$^{2+}$
to Eu$^{3+}$. However, the mean valence in them gets stabilized around
+2.3 at 10 GPa and +2.25 at 12.6 GPa, for EuFe$_{2}$As$_{2}$ and
EuCo$_{2}$As$_{2}$, respectively.\cite{Sun_11,Tan_16} Therefore,
the dip in $\mathit{T_{Eu}}(P)$ around 11 GPa in Fig. 6 is most likely
due to combined effects of the pressure-induced valence change of
Eu and the pressure-driven modification of the indirect Ruderman-Kittel-Kasuya-Yoshida
(RKKY) exchange interaction in Eu(Fe$_{0.925}$Co$_{0.075}$)$_{2}$As$_{2}$.
Since the RKKY exchange coupling depends strongly on the distance
between interlayer Eu$^{2+}$ moments, which is closely related to
the applied hydrostatic pressure, it is expectable that the magnetic
state of Eu as well as the ordering temperature, $\mathit{T_{Eu}}$,
will be tuned accordingly with increasing pressure. Unfortunately,
due to the limitation of the Paris-Edinburgh pressure cell used in
the neutron diffraction experiment, we can not achieve the pressure
above 11 GPa so as to conclude about the nature of the magnetic state
of Eu in this pressure region. The unobservable anomaly in the ac
susceptibility data at 13.2 GPa as shown in Fig. 2(c) tends to support
an antiferromagnetic order (either commensurate or incommensurate)
of the Eu sublattice. Therefore, we refer it as an ``unconfirmed
AFM state'' in Fig. 6. It is worth noting that a recent neutron diffraction
experiment has revealed an incommensurate antiferromagnetic in-plane
spiral order of the Eu$^{2+}$ spins as the ground-state magnetic
structure of Eu in EuCo$_{2}$As$_{2}$,\cite{Tan_16} the end member
of Eu(Fe$_{1-x}$Co$_{x}$)$_{2}$As$_{2}$ system. Combining the
phase diagram of the Eu magnetic ordering in Eu(Fe$_{1-x}$Co$_{x}$)$_{2}$As$_{2}$
established for the compositions with relatively lower Co concentrations
($\mathit{x}$ $\leq$ 18 \%),\cite{Jin_EuCoPhaseDiagram} it can
be concluded that with increasing Co doping level, the magnetic ground
state of Eu can be tuned from the in-plane commensurate AFM order
for $\mathit{x}$ = 0, via the intermediate canted AFM strcture, to
a pure FM order along the $\mathit{c}$ axis for $\mathit{x}$ $\approx$
18\%, then probably via some intermediate complex magnetic structure,
finally to the in-plane AFM spiral order for $\mathit{x}$ = 1. Co
doping drives the rotation of the Eu$^{2+}$ moments from the $\mathit{ab}$
plane to the $\mathit{c}$ axis, and then from the $\mathit{c}$ axis
back to the $\mathit{ab}$ plane. If we assume that the hydrostatic
pressure plays a similar role on the FM state of Eu as the electron
doping does, we might expect a ``novel'' magnetic state of the Eu$^{2+}$
spins for the pressure above 11 GPa, such as an incommensurate AFM
spiral order. Since we can not determine directly the nature of this possible ``novel'' magnetic state of Eu,
we refer it as an "unconfirmed AFM state'' in Fig. 6.

In addition, as hinted by the macroscopic measurements, the signature
of superconductivity emerges around 2.0 GPa. Therefore, the strong
ferromagnetic order of Eu at 2.0 GPa is compatible with the pressure-induced
superconductivity for Eu(Fe$_{0.925}$Co$_{0.075}$)$_{2}$As$_{2}$,
resembling the well confirmed coexistence of Eu-FM and the doping-induced
SC in several families of doped EuFe$_{2}$As$_{2}$.\cite{Ren_09,Jiao_11,Jiao_13,Jin_13,Jin_15,Jin_EuCoPhaseDiagram,Nandi_14,Nandi_14_neutron}
The pressure-induced superconducting dome of Eu(Fe$_{0.925}$Co$_{0.075}$)$_{2}$As$_{2}$,
however, is very narrow, similar to that of the parent compound EuFe$_{2}$As$_{2}$.\cite{Matsubayshi_11}

In conclusion, the effects of hydrostatic pressure on the static magnetism in Eu(Fe$_{0.925}$Co$_{0.075}$)$_{2}$As$_{2}$ are investigated by complementary electrical resistivity, ac magnetic susceptibility and single-crystal neutron diffraction measurements. A specific pressure-temperature phase diagram of Eu(Fe$_{0.925}$Co$_{0.075}$)$_{2}$As$_{2}$ is established. The structural phase transition, as well as the spin-density-wave order of Fe sublattice, is suppressed gradually with increasing pressure and disappears completely above 2.0 GPa. In contrast, the magnetic order of Eu sublattice persists over the whole investigated pressure range up to 14 GPa, yet displaying a non-monotonic variation with pressure. With the increase of the hydrostatic pressure, the magnetic state of Eu evolves from the canted antiferromagnetic structure in the ground state, via a pure ferromagnetic structure under the intermediate pressure, finally to a possible "novel'' antiferromagnetic structure under the high pressure. The strong ferromagnetism of Eu coexists with the pressure-induced superconductivity around 2 GPa. The change of the magnetic state of Eu in Eu(Fe$_{0.925}$Co$_{0.075}$)$_{2}$As$_{2}$ upon the application of hydrostatic pressure probably arises from the modification of the indirect Ruderman-Kittel-Kasuya-Yosida (RKKY) interaction between the Eu$^{2+}$ moments tuned by external pressure.

\bibliographystyle{apsrev} \bibliographystyle{apsrev}
\begin{acknowledgments}
This work is partly based on experiments performed on the D23 diffractometer
at the Institut Laue-Langevin (ILL), Gernoble, France. W. T. J. would
like to acknowledge S. Mayr for the assistance with the cutting and
orientation of the single crystal, and S. Klotz, C. Payre, E. Ressouche,
W. Schmidt and T. Hansen for their help with the high-pressure neutron
diffraction measurements. Z.B. acknowledges financial support from
the National Science Center of Poland, Grant No. 2011/01/B/ST5/06397.
JGC is supported by the National Science Foundation of China (Grant
No. 11574377), the National Basic Research Program of China (Grant
No. 2014CB921500), the Strategic Priority Research Program and the Key
Research Program of Frontier Sciences of the Chinese Academy of Sciences
(Grant Nos. XDB07020100, QYZDB-SSW-SLH013). Z. G. gratefully acknowledges
the financial support by the Swiss National Science Foundation (SNF
fellowship P2ZHP2-161980).

\bibliographystyle{apsrev}
\bibliography{EuCo_Pressure}

\begin{thebibliography}{10}
\bibitem{Footnote}
The ambient-pressure $\rho(T)$ curve of Eu(Fe$_{0.925}$Co$_{0.075}$)$_{2}$As$_{2}$ in Fig. 1(a) was measured on a small strip cut from the crystal and does not show a drastic drop below 11.4 K, indicating the filamentary nature of the superconducting-like behavior observed on a much larger crystal (Fig. 1 in Ref. 24).
\end{thebibliography}
\end{acknowledgments}

\end{document}